\documentclass[]{spie}  %>>> use for US letter paper

\usepackage{amsmath,amsfonts,amssymb}
\usepackage{graphicx}
\usepackage[colorlinks=true, allcolors=blue]{hyperref}
\usepackage{booktabs}
\usepackage{xspace}

\newcommand{\ie}{\textit{i.e.}\xspace}
\newcommand{\sref}{Sec.~\ref}
\newcommand{\tref}{Tab.~\ref}
\newcommand{\fref}{Fig.~\ref}

\title{ALF: an asymmetric Lyot wavefront sensor for the ELT/METIS vortex coronagraph}

\author[a]{G. Orban de Xivry}
\author[a]{O. Absil}
\author[a]{C. Delacroix}
\author[a,b]{P. Pathak}
\author[a]{M. Quesnel}
\author[c]{T. Bertram}

\affil[a]{Space sciences, Technologies, and Astrophysics Research (STAR) Institute, Université de Liège, allée du Six Août 19c, 4000 Liège, Belgium}
\affil[b]{Indian Institutes of Technology, Kanpur, India}
\affil[c]{Max-Planck-Institut für Astronomie, Königstuhl 17, 69117 Heidelberg, Germany}

\authorinfo{Further author information: (Send correspondence to Gilles Orban de Xivry)\\G. Orban de Xivry: E-mail: gorban@uliege.be.}

\begin{document} 
\maketitle

\begin{abstract}
Non-common path quasi-static and differential aberrations are one of the big hurdles of direct imaging for current and future high-contrast imaging instruments. They increase speckle and photon noise thus reducing the achievable contrast and lead to a significant hit in HCI performance.
The Mid-infrared ELT Imager and Spectrograph (METIS) will provide high-contrast imaging, including vortex coronagraphy in L, M and N bands, with the ultimate goal of directly imaging temperate rocky planets around the nearest stars.
Ground-based mid-infrared observations are however also impacted by water vapor inhomogeneities in the atmosphere, which generate additional chromatic turbulence not corrected by the near-infrared adaptive optics. This additional source of wavefront error (WFE) significantly impacts HCI performance, and even dominates the WFE budget in N band. Instantaneous focal plane wavefront sensing is thus required to mitigate its impact. In this context, we propose to implement a novel wavefront sensing approach for the vortex coronagraph using an asymmetric Lyot stop and machine learning. The asymmetric pupil stop allows for the problem to become solvable, lifting the ambiguity on the sign of even Zernike modes. Choosing the Lyot plane instead of the entrance pupil for this mask is also not arbitrary: it preserves the rejection efficiency of the coronagraph and minimizes the impact of the asymmetry on the throughput. Last but not least, machine learning allows us to solve this inversion problem which is non-linear and lacks an analytical solution.
In this contribution, 
we present our concept, our simulation framework, our results and a first laboratory demonstration of the technique. 

\end{abstract}


%%%%%%%%%%%%%%%%%%%%%%%%%%%%%%%%%%
\section{INTRODUCTION}
\label{sec:intro}  % \label{} allows reference to this section

The Mid-infrared ELT imager and spectrograph (METIS) is a first generation instrument of the Extremely Large Telescope currently under construction at Cerro Armazones. It will provide L, M, N band imaging and high dispersion spectroscopy, with high-contrast capabilities. METIS recently passed its final design review. It is  now well in its construction phase, and is expected to see its first light towards the end of the decade\cite{Brandl+24}.

Among a wide array of science topics, METIS is expected to bring major breakthroughs in exoplanet detection and characterisation, and in our understanding of protoplanetary disks. To this end, METIS is specifically designed with high-contrast imaging capabilities in mind, including a high-performance single-conjugated adaptive optics module (see e.g. Bertram et al.\cite{Bertram+24}) and several advanced coronagraphic concepts (see e.g. Absil et al.\cite{Absil+24}): vortex coronagraphs to reach the smallest inner working angle and high throughput, and grating-vector apodising phase plates implemented for more robustness to aberrations.

The HCI performance is influenced by a large range of instrumental and environmental effects, which has been the subject of detailed analyses\cite{Carlomagno+20, Delacroix+22}. Overall, non-common path aberrations (NCPA) are expected to play a major role in the overall performance budget, and our abilitiy to measure and correct them will be a key driver for the final HCI performance.
In the framework of METIS, we can distinguish two main time-varying NCPA contributors: i) quasi-static aberrations essentially due to slight variation of the beam footprint on the optics, ii) dynamical aberration caused by water vapor in the atmosphere. Indeed water vapor becomes increasingly dispersive at infrared wavelengths leading to differential aberrations between the METIS SCAO working at K band and the science channel working at L, M or N band. Specifically, we expect about 300nm rms wavefront error due to water vapour seeing at 11.5$\mu$m. We refer to Absil et al.\cite{Absil+22} for thorough discussion of water vapour seeing effect and how it is modeled for METIS.
Because we expect water vapour to be displaced by wind, similarly to the dry air seeing, this contribution is highly dynamic and requires an instantaneous focal plane wavefront sensing approach to effectively mitigate it.

In this contribution, we propose to implement a novel wavefront sensing approach for the vortex coronagraph of METIS using an asymmetric Lyot stop and machine learning. In \sref{sec:ALF}, we shortly introduce our concept. In \sref{sec:simu}, we present the simulation framework we have implemented for this work. In \sref{sec:results}, we present several simulation results for classical imaging, vortex coronagraph, and our first laboratory results. We conclude in \sref{sec:conclusion} with our perspectives.

%%%%%%%%%%%%%%%%%%%%%%%%%%%%%%%%%%
\section{ASYMMETRIC LYOT MASK WAVEFRONT SENSOR}\label{sec:ALF}
Martinache et al.\cite{Martinache+13}  first proposed the asymmetric pupil wavefront sensor (APF-WFS) method for classical imaging. It relies
on the Fourier properties of the images acquired after an asymmetric mask has been placed in the entrance pupil. In particular, he showed that, in the low-aberration regime, the phase $\Phi$ in the
Fourier plane (or OTF plane) of an image can be related linearly to the entrance pupil phase $\phi$. Specifically, for a point source, we have
\begin{equation}
    \Phi = \mathbf{A} \phi,
\end{equation}
where $\mathbf{A}$ corresponds to the phase transfer matrix. This matrix establishes the mapping between the discretized representations of the two spaces and is thus generally rectangular. By pseudo-inversion (whose existence is ensured by the presence of the asymmetry in the pupil), the pupil phase can  be derived.

This linear formalism is only
valid for normal imaging and cannot be applied to focal plane coronagraphs which, by suppressing the on-axis light, break the shift-invariance properties implicitly required by this Fourier approach\cite{Laugier2020}
Nevertheless an asymmetric (pupil or Lyot) mask, by breaking the centro-symmetry, should make the problem solvable, lifting the ambiguity on the sign of even Zernike modes. What is missing is thus a solution to this non-linear problem, \ie retrieving the entrance pupil phase aberration from a focal plane post-coronagraphic image. Since we lack an analytical solution, we resort to deep learning to determine one, and more specifically use convolutional neural networks (CNNs) in a supervised learning approach.

For the vortex coronagraphic mode, we choose the Lyot plane instead of the entrance pupil to insert the asymmetric mask. It allows us to preserves the coronagraphic rejection efficiency and to minimise the impact of the asymmetry on the performance.
Hence, our proposed concept is named the Asymmetric Lyot mask waveFront sensor, or ALF.

%%%%%%%%%%%%%%%%%%%%%%%%%%%%%%%%%%
\section{SIMULATION FRAMEWORK}\label{sec:simu}

In the context of the METIS final design review, %FDR, 
we have developed a simulation framework that includes system-level knowledge such as the instrument mode, the SCAO residuals, the water vapour seeing, and the quasi-static NCPAs. The framework is coded in Python and uses the HCIPy\cite{Por+18} package to define the optical model of the instrument and to perform optical propagations.

The framework is based essentially on two types of classes: the sensor on one hand, and  the instrument on the other hand. 
The sensors we have implemented so far include  the kernel (or APF-WFS), a generic deep learning sensor (particularly for ALF), and phase sorting interferometry (PSI, see, for example, Codona et al. \cite{Codona+13}). In this contribution, we focus on the first two sensors, \ie using the kernel approach or  deep learning.

\begin{figure}[ht]
    \centering
    \includegraphics[width=1\linewidth]{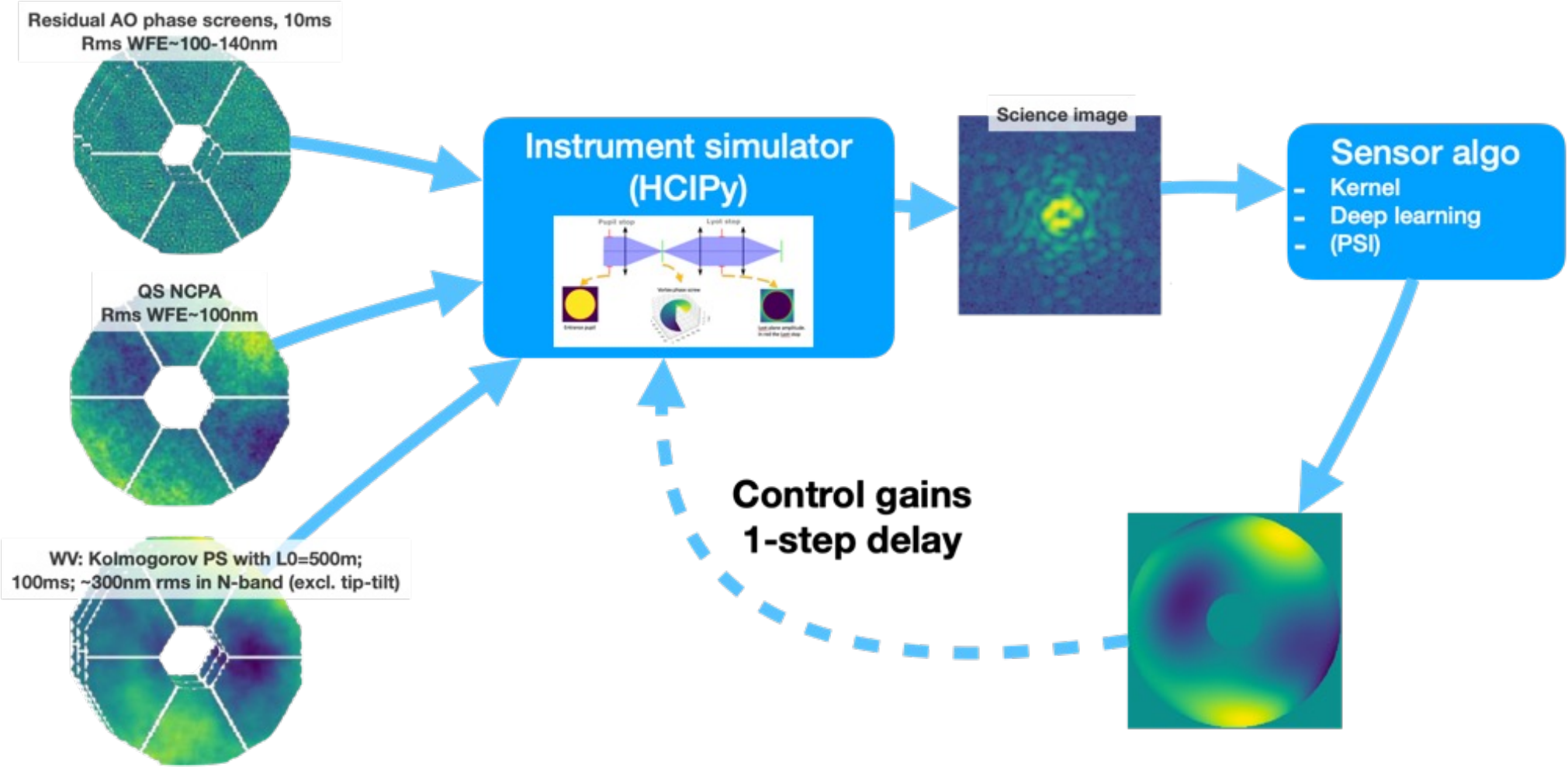}
    \caption{Sketch of our simulation framework. Our simulations take a series of phase screen buffers as input. An instrument simulator allows to simulate specific observing mode with ELT/METIS, such as classical imaging or classical vortex coronagraphy. The produced science images are then fed to a sensor which will derive a phase map used as a correction in closed-loop.}
    \label{fig:framework}
\end{figure}

Our framework is illustrated in \fref{fig:framework}. The adaptive optics residuals and water vapour seeing phase screens are generated off-line using the METIS-AO simulator based on COMPASS\cite{Gratadour+16}. The quasi-static NCPA, due to chromatic beam wander, were obtained via a dedicated modeling strategy of the atmospheric refraction using ZEMAX\cite{Bone+}

%%%%%%%%%%%%%%%%%%%%%%%%%%%%%%%%%%
\section{RESULTS \& PERFORMANCE}\label{sec:results}
We focus our simulations on the N band imaging case, where the impact of water vapour seeing is most severe.
First, we investigate the normal imaging case, exploring the impact of the asymmetric mask configuration on  performance and establishing a global error budget.
Next, we turn our attention to the vortex coronagraph, exploring different mask configurations, and performing a representative closed-loop simulation.
Finally, we present our initial attempt to experimentally test ALF on our VODCA bench.

\subsection{Normal imaging}
For normal imaging, we use the kernel formalism, see \sref{sec:ALF}, and rely on the XARA package\cite{Martinache+13}\footnote{see also \texttt{https://github.com/fmartinache/xara/}} which we integrate to our simulation framework.

\begin{figure}[ht]
    % \centering
    \includegraphics[width=0.5\linewidth]{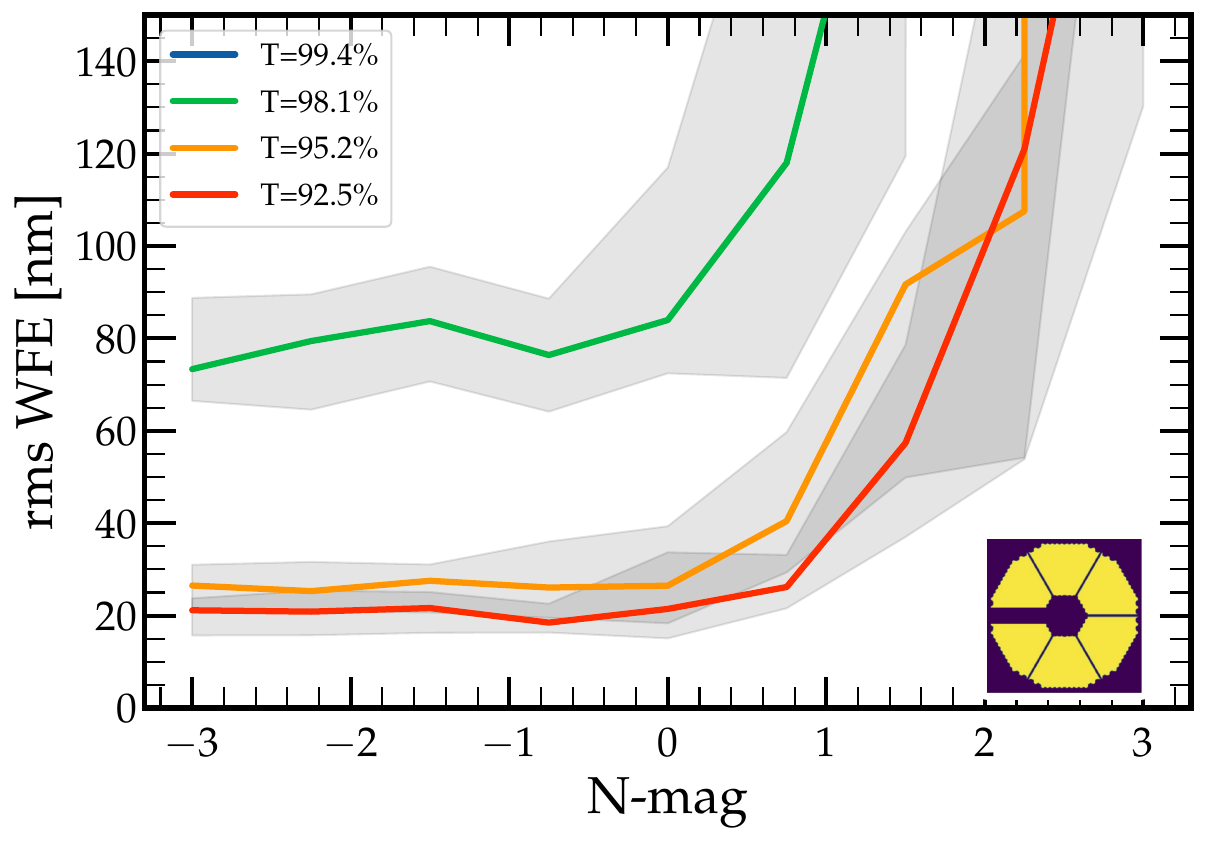}
    \includegraphics[width=0.5\linewidth]{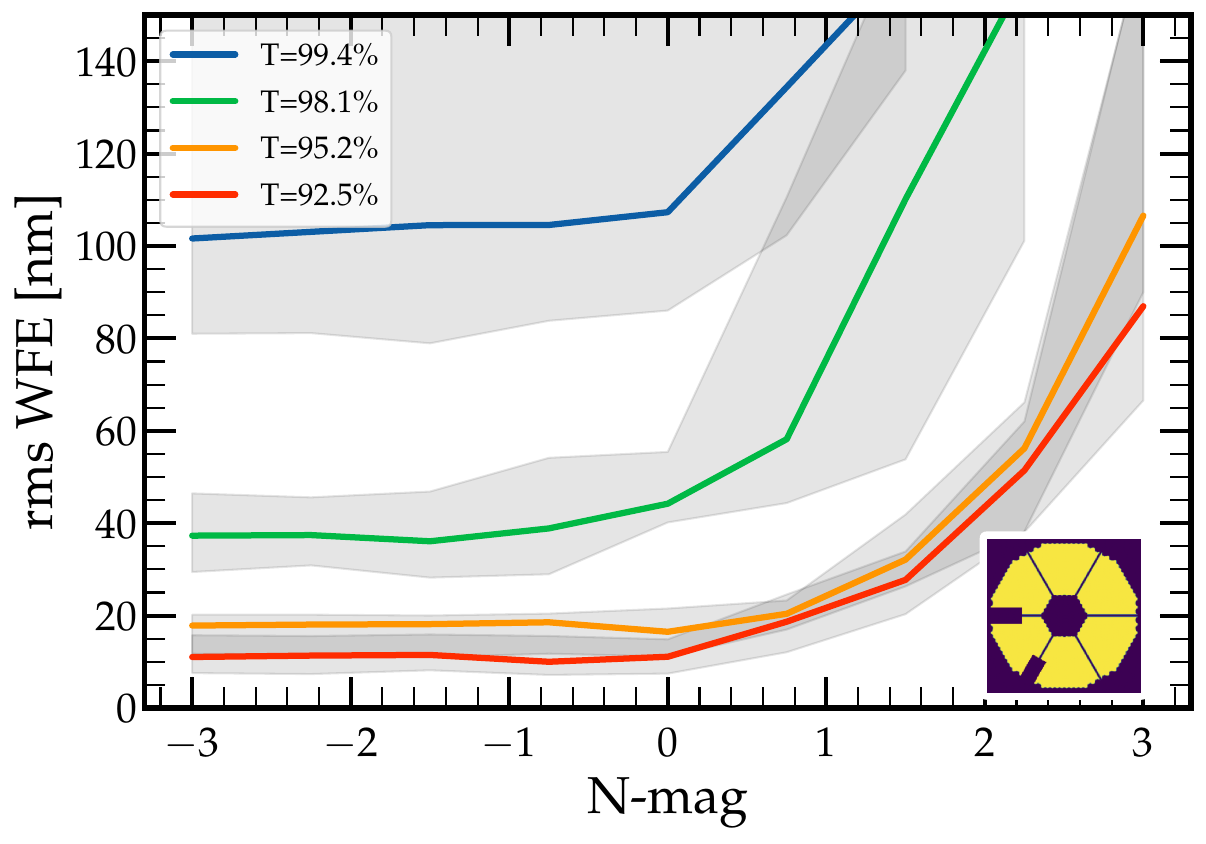}
    \caption{Residual rms wavefront error as a function of N-band magnitude. The WFE is measured on the controlled 20 Zernike modes. The different colors corresponds to different asymmetric spider thicknesses of 2.5, 5, 10, and 15\% of the pupil diameter. (Left) one asymmetric spider. (Right) two half spiders.}
    \label{fig:IMG}
\end{figure}

We compare two asymmetric mask configurations: one with a thicker spider, and one with two thicker half-spiders. For those two configurations, we explore different  thicknesses. The sensing performance is evaluated for different N band magnitudes on static aberrations. The results are shown in \fref{fig:IMG}. The figure illustrates the impact of the mask configuration and the thickness of the asymmetry.
The best results are obtained for a mask with two half spiders of thicknesses $\geq$ 10\% of the pupil diameter, leading to a reduction of transmission of $\geq$4.8\%.

We also analyse the impact of polychromaticity and the response in closed-loop, allowing us to derive a global error budget, see \tref{tab:budget}. The fitting and temporal errors are analyzed analytically, as described in the following two paragraphs.

Assuming phase Kolmogorov spectrum with pre-corrected tip-tilt, the modal fitting error can be approximated by\cite{Noll76} $\sigma_{\rm fitting}^2 \sim 0.134 \left(0.2944 N_{modes}^{- \sqrt{3}/2} (D/r_0^{wv})^{5/3} \right)$. Assuming an equivalent characteristics scale $r_0^{wv}=95m$ for water vapor (equivalent to 300nm rms of variance, excluding tip-tilt), the residual rms fitting error for 20 corrected Zernike modes amounts to about 120nm rms and about 60nm rms for 100 corrected modes.

The temporal error can be estimated with $\sigma_{temp}^2 = \left(f_G^{wv} / f_{BW}\right)^2$. The Greenwood frequency is given by $f_G^{wv} = 0.427 V_{wind} /r_0^{wv} \sim 0.04$~Hz assuming an equivalent wind speed of $V_{wind}\sim 8.8$~m/s.  The bandwidth frequency $f_{BW}$ is the sensor framerate divided by $ \sim 12$, following standard practices. As we are only interested in the variance of the first 22 modes excluding tip-tilt,  we scale the temporal variance by a factor\cite{Noll76} $(\Delta_3 - \Delta_{22}) \sim 0.11$. Altogether, we obtain a temporal error of about 50nm rms for a framerate of 10~Hz and $>$300~nm rms at 1~Hz. Note that the temporal error estimation at 1~Hz is likely overestimated since the analytical formula (the definition of the Greenwood frequency in particular) assumes a constant power law and ignores the low-frequency breaks due to the removal of average phase\cite{Conan+95}.

\begin{table}[ht]
\caption{Focal plane wavefront error budget at N band for normal imaging. The first column provides our baseline, correcting 20 Zernike modes and running at 10~Hz. The second column provides variations to the baseline, by increasing the number of modes (impacting the sensor noise and the fitting error) and decreasing the frame rate (impacting the temporal error).}
\label{tab:budget}
\begin{center}
\begin{tabular}{lll}
\toprule
Error term & Baseline [20 modes, 10~Hz] & [100modes, 10~Hz] \\
\midrule
 Sensor noise     & 20  &  50   \\
 Chromatic error (20\% bandwidth) & 48  &  48  \\
 Fitting error    & 120 & 60   \\
 Temporal error   & 50  & $\sim$300\\
 \bottomrule
 Total [nm rms]   & 140 & 330 \\
 \bottomrule
\end{tabular}
\end{center}
\end{table}

\subsection{Classical vortex coronagraph}
Since we lack an analytical solution, we use a deep learning approach to retrieve the phase from focal plane images. We implement a supervised learning approach with a workflow as follows: 
\begin{enumerate}
    \item we generate a large labelled dataset consisting of science images and the corresponding Zernike coefficients,
    \item we train a neural network using a typical 90\%-10\% split for training and validation,
    \item we use our trained model for evaluation on a different dataset.
\end{enumerate}

For the neural network architecture, we use a ResNet\cite{He+15} with 18 layers. Our typical training datasets are based on 10,000 random phase screens.

\begin{figure}[ht]
    \centering
    \includegraphics[width=0.7\linewidth]{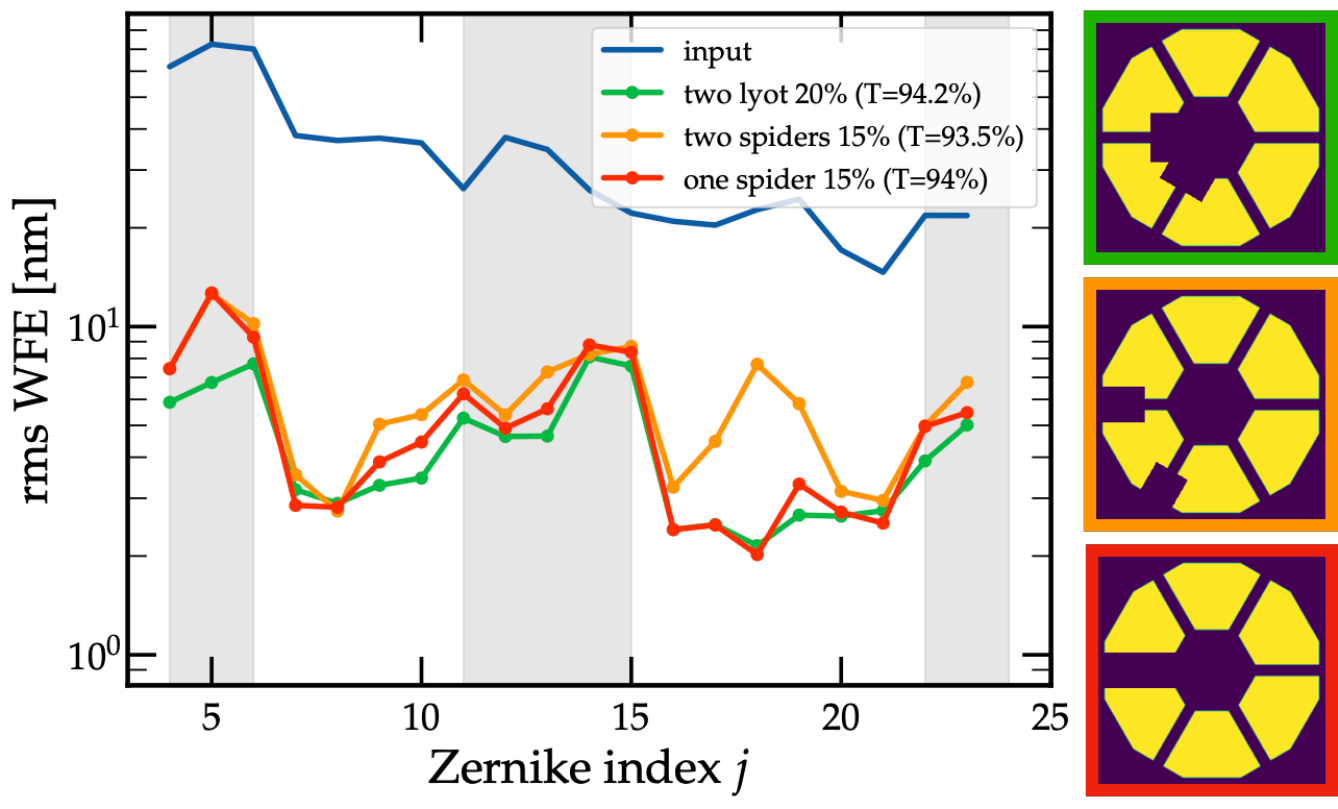}
    \caption{Modal rms wavefront error for different Lyot stop configurations. }
    \label{fig:CVC_modal}
\end{figure}

Once again, we compare different mask configurations. On top of the two masks used for classical imaging, we also test a mask with two half spiders starting from the central obscuration rather than the external diameter: with a vortex coronagraph, the central obscuration impacts the light scattered in the downstream Lyot plane which appears as a monotonically decreasing flux moving away from the central obscuration. Hence we expect more flux in this region and by placing an asymmetry there, we might obtain a stronger signature in the focal plane.
The three masks are compared in \fref{fig:CVC_modal} by decomposing the residual wavefront error on the Zernike modal basis. We observe typically a larger error on even modes (which would not be sensed without asymmetry), and  the mask configuration significantly impacts  the sensing. The mask with two  half spiders starting from the central obscuration perform best.

\begin{figure}[ht]
    \centering
    \includegraphics[width=0.7\linewidth]{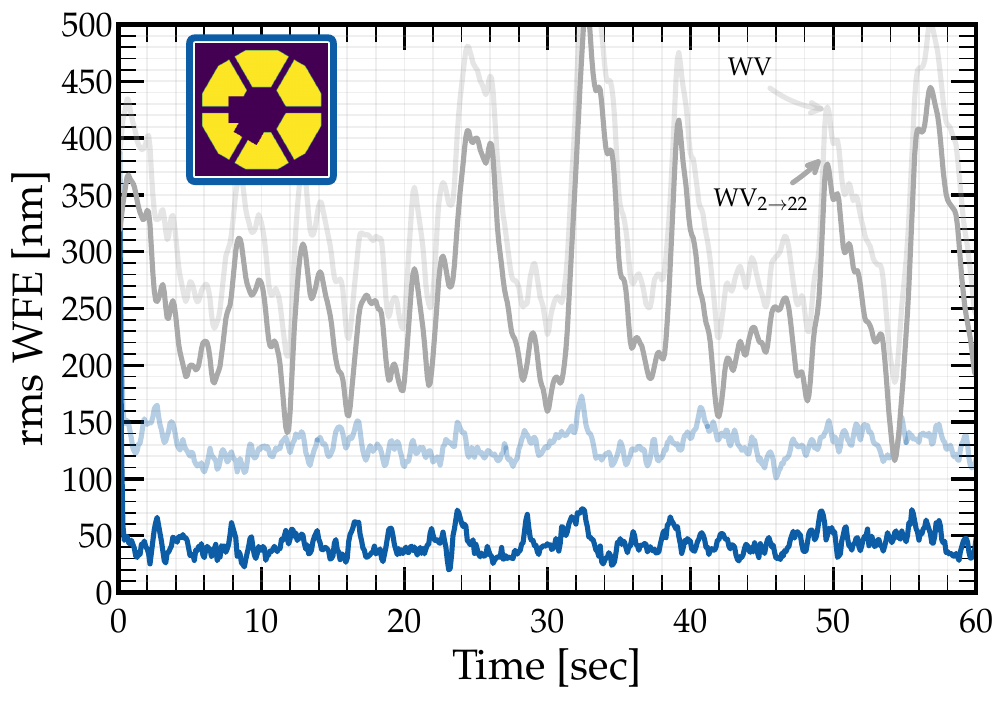}
    \caption{Closed-loop simulation in the presence of atmospheric residuals, with 0.1s DIT time. The rms WFE on the first 20 modes reduced from $\sim$260nm to $<$50nm.}
    \label{fig:CVC_closed-loop}
\end{figure}

Finally, using our best neural network model, we test our approach in closed-loop, \ie injecting the residual AO phase screens and the variable water vapor seeing. This closed-loop simulation is a 1-min simulation with science DIT of 100msec, residual AO phase screens  sampled every 10 msec (10 realisations per science image), and water vapour seeing phase screens sampled every 100msec (1 realisation per science image). We use the expected METIS N band photometry, and produce realistic classical vortex images with a resolution of 9 px/$\lambda/D$, a field-of-view of $5 \lambda/D$, and a pupil plane resolution of 256 pixels.
The results are shown in \fref{fig:CVC_closed-loop}. The input water vapour seeing is about 330nm rms, with 270 nm rms on the first 20 modes. With ALF, we obtain a residual rms wavefront error of about 140nm rms on all modes, and below 50nm rms on the first 20 modes. The 140nm rms is on par with our error budget \tref{tab:budget} and suggests that we reach similar performance to normal imaging.

\subsection{First laboratory tests}
To validate our concept experimentally, we use our mid-infrared coronagraphic VODCA bench\cite{Jolivet+19}.
It features a vortex coronagraph and an ALPAO deformable mirror with 97 actuators in the pupil plane.
For this first test, we use circular apertures and insert an asymmetric Lyot stop with a single thick bar.
We acquire large datasets (from 1,000 to 10,000 different realisations), and train our ResNet-18 on those experimental data. Finally, we test the trained model by injecting manually aberrations with our DM. The model retrieves correctly the injected aberrations, and works also well when applied iteratively. At this stage, we have not yet precisely quantified  the performance of our approach. Our current focus is to demonstrate that the sensor works for the correct reasons, \ie, to  explicitly show that  sensing is enabled by the asymmetry and not due to bench imperfections, such as imperfect DM calibration or non-uniform pupil illumination.

%%%%%%%%%%%%%%%%%%%%%%%%%%%%%%%%%%
\section{CONCLUSION \& PERSPECTIVES}\label{sec:conclusion}

In this contribution, we have presented a new concept for performing focal plane wavefront sensing with the vortex coronagraph, \ie, ALF, the asymmetric Lyot plane wavefront sensor.
We have demonstrated a proof-of-concept by implementing a full simulation framework representative of the METIS HCI modes. Specifically, we have explored the impact of the asymmetric mask configuration on performance, established an error budget for our sensor, and  evaluated  performance in closed-loop.

Future work will focus on

\begin{itemize}
    \item Further optimizing the asymmetric Lyot mask and the number of corrected modes.
    \item Analyzing other HCI modes beyond the classical vortex coronagraph, such as  the ring-apodizer vortex coronagraph. 
    \item Re-evaluating the global HCI performance in terms of post-processed contrast,  using the achieved residual error after focal plane wavefront sensing and correction.
    \item Investigating more advanced predictive control to boost the rejection bandwidth of our sensor. 
    \item Improving our laboratory tests with better bench calibration and using ELT-like pupils to obtain more representative results.
    \item Exploring possibilities to perform an on-sky demonstration on a 8-m class telescope.
\end{itemize}

%%%%%%%%%%%%%%%%%%%%%%%%%%%%%%%%%%
\acknowledgments % equivalent to \section*{ACKNOWLEDGMENTS}       
This project has received funding from the European Research Council (ERC) under the European Union’s Horizon 2020 research and innovation programme (grant agreement No 819155), and from the Wallonia-Brussels Federation (grant for Concerted Research Actions).

% References
\bibliography{report} % bibliography data in report.bib
\bibliographystyle{spiebib} % makes bibtex use spiebib.bst

\end{document}